\documentclass[12pt,number,preprint,times,letterpaper]{elsarticle}
\usepackage{algorithm}
\usepackage{algorithmic} 
\usepackage{amssymb}
\usepackage{amsthm}
\usepackage{amsmath}
\usepackage{mathtools}
\usepackage{mathrsfs}
\usepackage{dsfont}

\usepackage{subfig}

\usepackage{breakcites}

\usepackage[
  hidelinks,
	breaklinks=true,
]{hyperref}
\hypersetup{
  pdfauthor={\ldots},
  % ...
}
\usepackage{bookmark}

\newdefinition{rmk}{Remark}
\newdefinition{definition}[thm]{Definition}
\newproof{pf}{Proof}
\newproof{pot}{Proof of Theorem \ref{thm2}}

\biboptions{authoryear}

\journal{Economics Letters, \textbf{Accepted on Nov. 24}}

\begin{document}

\begin{frontmatter}
	\title{Wasserstein Index Generation Model:\\
		Automatic Generation of Time-series Index with Application to
		Economic Policy Uncertainty}

	\author{Fangzhou Xie\fnref{l1}} 
	\ead{fangzhou.xie@nyu.edu}

	\fntext[l1]{Present Mailing Address:
		546 Main St, Apt 437, New York, NY, 10044.}
	\address{Department of Economics, New York University}

	\date{\today}

	\begin{abstract}
		I propose a novel method,
		the Wasserstein Index Generation model (WIG),
		to generate a public sentiment index automatically.
		To test the model's effectiveness,
		an application to generate Economic Policy Uncertainty (EPU) index
		is showcased.
	\end{abstract}



\end{frontmatter}

\section{Introduction}\label{sec:intro}

\cite{baker2016} has created a novel method to measure Economic Policy
Uncertainty, the EPU index,
which has attracted significant attention and been followed by a strand of literature
since its proposal.
However, it entails a carefully designed framework and significant manual efforts
to complete its calculation.
Recently, there has been significant progress in the methodology
of the generation process of EPU\@, e.g.\
differentiating contexts for uncertainty~\citep{saltzman2018},
generating index based on Google Trend~\citep{castelnuovo2017},
and correcting EPU for Spain~\citep{ghirelli2019}.
I wish to extend the scope of index-generation
by proposing this generalized method,
namely the Wasserstein Index Generation model (WIG).

This model (WIG)
incorporates several methods that are widely used in machine learning,
word embedding~\citep{mikolov2013},
Wasserstein Dictionary Learning~\citep[WDL]{schmitz2018},
Adam algorithm~\citep{kingma2015},
and Singular Value Decomposition (SVD).
The ideas behind these methods are essentially dimension reduction.
Indeed, WDL reduces the dimension of the dataset into its bases
and associated weights, and SVD could shrink the dimension of
bases once again to produce unidimensional indices for further
analysis.

I test WIG’s effectiveness in generating the
Economic Policy Uncertainty index~\citep[EPU]{baker2016},
and compare the result against existing ones~\citep{azqueta-gavaldon2017},
generated by the auto-labeling Latent Dirichlet Allocation
\citep[LDA]{blei2003} method.
Results reveal that
this model requires a much smaller dataset to achieve better results,
without human intervention.
Thus, it can also be applied for generating other time-series
indices from news headlines in a faster and more efficient manner.

Recently, \cite{shiller2017} has called for more attention in collecting
and analyzing text data of economic interest.
The WIG model responds to
this call in terms of generating time-series sentiment
indices from texts by facilitating machine learning algorithms.

\section{Methods and Material}
\subsection{Wasserstein Index Generation Model}\label{subsec:wig}
\citet{schmitz2018} proposes an unsupervised machine learning technique to
cluster documents into topics, called the Wasserstein Dictionary Learning (WDL),
wherein both documents and topics are considered as discrete
distributions of vocabulary.
These discrete distributions can be reduced into bases and corresponding
weights to capture most information in the dataset and thus
shrink its dimension.

Consider a corpus with \(M\) documents and a vocabulary of \(N\) words.
These documents form a matrix of
\(Y=\left[y_{m}\right] \in \mathbb{R}^{N \times M}\),
where \(m \in\left\{1, \dots, M\right\}\),
and each \(y_{m} \in \Sigma^{N}\).
We wish to find topics \(T \in \mathbb{R}^{N \times K}\),
with associated weights \(\Lambda \in \mathbb{R}^{K \times M}\).

In other words, each document is a discrete distribution,
which lies in an \(N\)-dimensional simplex.
Our aim is to represent and reconstruct these documents according to some
topics \(T \in \mathbb{R}^{N \times K}\), with associated weights
\(\Lambda \in \mathbb{R}^{K \times M}\), where
\(K\) is the total number of topics to be clustered.
Note that each topic is a distribution of vocabulary,
and each weight represents its associated document as a weighted barycenter
of underlying topics.
We could also obtain a distance matrix of the total vocabulary
\(C^{N \times N}\), by first generating word embedding
and measuring word distance pairwise by using a metric function,
that is, \(C_{ij} = d^2(x_i, x_j)\), where
\(x \in \mathbb{R}^{N \times D}\),
\(d(\boldsymbol{\cdot})\) is Euclidean distance,
and \(D\) is the embedding depth.
\footnote{\cite{saltzman2018}
	proposes differentiating the use of ``uncertainty'' in both
	positive and negative contexts.
	In fact, word embedding methods, for example, \ Word2Vec \citep{mikolov2013},
	can do more. They consider not only the positive
	and negative context for a given word, but
	all possible contexts for all words.
}

Further, we could calculate the distances between documents and
topics, namely the Sinkhorn distance.
It is essentially a \(2\)-Wasserstein distance,
with the addition of an entropic regularization
term to ensure faster computation.
\footnote{One could refer to \cite{cuturi2013} for the Sinkhorn algorithm
	and \cite{villani2003} for the theoretic results in optimal transport.}

\begin{definition}[Sinkhorn Distance]
	Given \(\mu, \nu \in \mathscr{P}(\Omega)\),
	\(\mathscr{P}(\Omega)\) as a Borel probability measure on \(\Omega\),
	\(\Omega \subset \mathbb{R}^{N}\), and \(C\) as cost matrix,
	\begin{equation}\label{def:sinkhorn}
		\begin{aligned}
			S_{\varepsilon} (\mu, \nu; C) & :=  \min_{\pi \in
				\Pi(\mu, \nu)}  \langle\pi , C\rangle + \varepsilon \mathcal{H}(\pi) \\
			s.t.\ \Pi(\mu, \nu)           & :=\left\{\pi \in \mathbb{R}_{+}^{N
				\times N}, \pi \mathds{1}_{N}=\mu, \pi^{\top}
			\mathds{1}_{N}=\nu\right\},
		\end{aligned}
	\end{equation}
	where \(\mathcal{H}(\pi) := \langle\pi,\log(\pi)\rangle\)
	and \(\varepsilon\) is Sinkhorn weight.
\end{definition}

Given the distance function for a single document,
we could set up the loss function for the training process:

\begin{equation}\label{eq:lossfcn}
	\begin{aligned}
		 & \min_{R, A} \sum_{m=1}^{M} \mathcal{L}\left(y_m, y_{S_{\varepsilon}}
		\left(T(R), \lambda_m(A) ; C, \varepsilon\right)\right),                \\
		 & given~~t_{nk}(R) := \frac{e^{r_{nk}}}{\sum_{n'} e^{r_{n'k}}},~~
		\lambda_{nk}(A) := \frac{e^{a_{km}}}{\sum_{k'} e^{a_{k'm}}}.
	\end{aligned}
\end{equation}

In Equation~\ref{eq:lossfcn},
\(y_{S_{\varepsilon}}\left(\boldsymbol{\cdot}\right)\)
is the reconstructed document given topics \(T\) and weight\(\lambda\)
under Sinkhorn distance (Equation.~\ref{def:sinkhorn}).
Moreover, the constraint that \(T\) and \(\Lambda\) being distributions
in Equation~\ref{def:sinkhorn} is automatically
fulfilled by column-wise \textit{Softmax} operation in the loss function.
The process is formulated in Algorithm~\ref{alg:wdl},
where we first initialized matrix \(R\) and \(A\) by taking a random sample
from a Standard Normal distribution and take \textit{Softmax}
on them to obtain \(T\) and \(\Lambda\).
\(\nabla_{T}\mathcal{L(\boldsymbol{\cdot}~;\varepsilon)}\) and
\(\nabla_{\Lambda}\mathcal{L(\boldsymbol{\cdot}~;\varepsilon)}\)
are the gradients taken from the loss function with respect to
topics \(T\) and weights \(\Lambda\).
The parameters \(R\) and \(A\) are then optimized
by the Adam optimizer with the gradient at hand and learning rate \(\rho\).
\textit{Softmax} operation is operated again to ensure
constraints being unit simplex (as shown in Equation~\ref{eq:lossfcn}).

\begin{algorithm}
	\caption{Wasserstein Index Generation\protect}
	\begin{algorithmic}[1]
		\REQUIRE Word distribution matrix \(Y\). Batch size \(s\).\\
		Sinkhorn weight \(\varepsilon\). Adam Learning rate \(\rho\).
		\ENSURE Topics \(T\), weights \(\Lambda\).
		\STATE Initialize \(R, A \sim \mathcal{N}(0,1)\).
		\STATE \(T \leftarrow Softmax(R)\), \(\Lambda \leftarrow Softmax(A)\).
		\FOR{Each batch of documents}
		\STATE
		\(R \leftarrow R - Adam\left(\nabla_{T}
		\mathcal{L(\boldsymbol{\cdot}~;\varepsilon)};~\rho\right)\),\\
		\(A \leftarrow A - Adam\left(\nabla_{\Lambda}
		\mathcal{L(\boldsymbol{\cdot}~;\varepsilon)};~\rho\right)\).
		\STATE \(T \leftarrow Softmax\left(R\right)\),
		\(\Lambda \leftarrow Softmax\left(A\right)\).
		\ENDFOR
	\end{algorithmic}
	\label{alg:wdl}
\end{algorithm}

Next, we generate the time-series index.
By facilitating Singular Value Decomposition (SVD) with one component,
we can shrink the dimension of vocabulary
from \(T^{N \times K}\) to \(\widehat T^{1 \times K}\).
Next, we multiply \(\widehat T\) by \(\Lambda^{K \times M}\)
to get \(Ind^{1 \times M}\), which is the document-wise score given
by SVD\@.
Adding up these scores by month and scaling the index to get a mean of
100 and unit standard deviation, we obtain the final index.

\subsection{Data and Computation}
I collected data from \textit{The New York Times}
comprising news headlines from Jan.~1, 1980 to Dec.~31, 2018.
The corpus contained 11,934 documents, and 8,802 unique tokens.
\footnote{
	Plots given in Figure~\ref{results}, however, are from
	Jan.~1, 1985 to Aug.~31, 2016 for maintaining the same range to be compared
	with that from~\cite{azqueta-gavaldon2017}.}

Next, I preprocess the corpus for further training process, for example,
by removing special symbols, combining entities,
and lemmatizing each token.
\footnote{Lemmatization refers to the process of converting each word to
	its dictionary form according to its context.}
Given this lemmatized corpus, I facilitate Word2Vec to generate embedding
vectors for the entire dictionary and thus am able to calculate the distance
matrix \(C\) for any pair for words.

To calculate the gradient (as shown in Algorithm~\ref{alg:wdl}),
I choose the automatic differentiation library,
PyTorch \citep{paszke2017}, to perform differentiation of the loss function
and then update the parameters using the Adam algorithm \citep{kingma2015}.

To determine several important hyper-parameters, I use cross validation
as is common in machine learning techniques.
One-third of the documents are set for testing data and the rest are used for
the training process:
Embedding depth \(D = 10\),
Sinkhorn weight \(\varepsilon = 0.1\),
batch size \(s = 64\),
topics \(K = 4\),
and Adam learning rate \(\rho = 0.005\).
Once the parameters are set at their optimal values,
the entire dataset is used for training, and thus, the topics
\(T\) and their associated weights \(\Lambda\) are obtained.

\section{Results}\label{results}

\begin{figure}[H]
	\centering 
	\includegraphics[width=\linewidth]{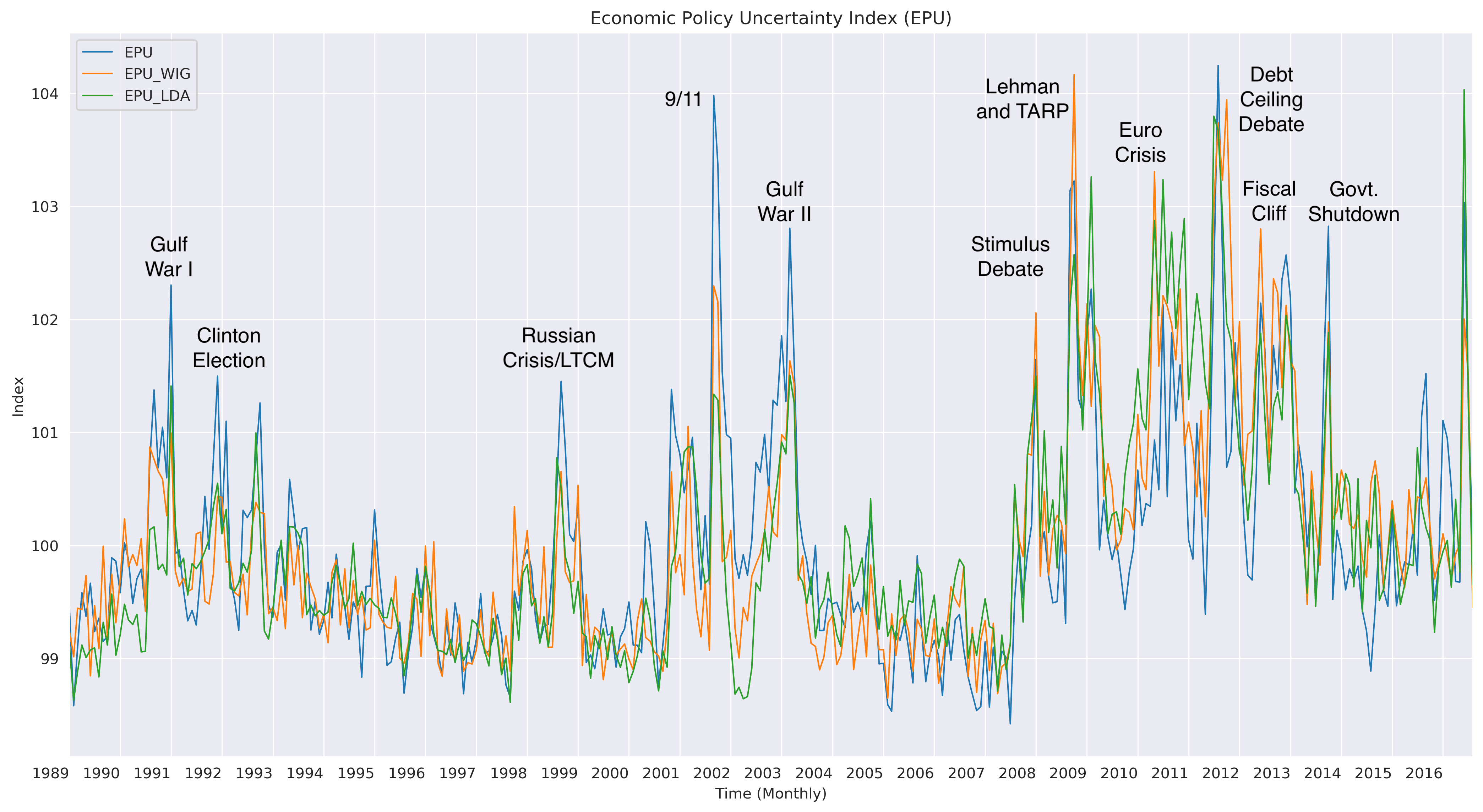}
	\caption{
		Original EPU~\protect\citep{baker2016},
		EPU with LDA~\protect\citep{azqueta-gavaldon2017},
		and EPU with WIG in Sec.~\protect\ref{subsec:wig}.
	}
	\label{fig:epu}
\end{figure}

As shown in Figure~\ref{fig:epu}, the EPU index generated by the WIG model
clearly resembles the original EPU\@.
Moreover, the WIG detects the emotional spikes better than LDA,
especially during major geopolitical events, such as
``Gulf War I,'' ``Bush Election,'' ``9/11,''
``Gulf War II,'' and so on.
For comparison, I calculated the cumulated difference between the original EPU
with that generated by WIG and LDA, respectively
(Figure~\ref{fig:cumsumdiff}, \ref{appen}).
Results indicate that the WIG model slightly out-performs LDA.

To further examine this point, I apply the Hodrick–Prescot filter\footnote{
	The HP filter was applied with a monthly weighted parameter 129600.
}
to three EPU indices,
and calculate the
Pearson's and Spearman's correlation factors between
the raw series, cycle components, and trend components,
as shown in~\ref{tab:correlation}, \ref{appen}.
These tests also suggest that the series generated by WIG
capture the EPU's behavior better than LDA over this three-decade period.

Moreover, this method only requires a small dataset compared with LDA\@.
The dataset used in this article contains only news headlines, and
the dimensionality of the dictionary is only a small fraction compared with
that of the LDA method. The WIG model takes only half an hour for computation
and still produces similar results.\footnote{
	Comparison of datasets are in Table~\ref{tab:comparison}, \ref{appen}.
}

Further, it extends the scope of automation in the generation process.
Previously, LDA was considered an automatic-labeling method,
but it continues to require human interpretation
of topic terms to produce time-series
indices. By introducing SVD, we could eliminate this requirement
and generate the index automatically as a black-box method.
However, it by no means loses its interpretability. The key terms are still
retrievable, given the result of WDL, if one wishes to view them.

Last, given its advantages,
the WIG model is not restricted to generating EPU, but could potentially
be used on any dataset regarding a certain topic
whose time-series sentiment index
is of economic interest. The only requirement is that the input corpus
be related to that topic, but this is easily satisfied.

\section{Conclusions}
I proposed a novel method to generate time-series indices of economic interest
using unsupervised machine learning techniques.
This could be applied as a black-box method, requiring only a small dataset,
and is applicable to any time-series indices' generation.
This method incorporates deeper methods from machine learning research,
including word embedding, Wasserstein Dictionary Learning,
and the widely used Adam algorithm.

\section*{Acknowledgements}
I am grateful to Alfred Galichon for launching this project and
to Andr\'es Azqueta-Gavald\'on for kindly providing his EPU data.
I would also like to express my gratitude to referees at the
3rd Workshop on Mechanism Design for Social Good (MD4SG~'19)
at ACM Conference on Economics and Computation (EC~’19)
and the participants at the Federated Computing Research Conference (FCRC 2019)
for their helpful remarks and discussions.
I also appreciate the helpful suggestions from the anonymous referee.

This research did not receive any specific grant from funding agencies
in the public, commercial, or not-for-profit sectors.

\bibliography{wigpaper}

\newpage
\appendix
\section{Supplementary Materials}\label{appen}

\begin{table}[H]
	\centering
	\begin{tabular}{llllll}
		\hline
		\textbf{Name} & \textbf{Method} & \textbf{Type} & \textbf{Num.~Entries} & \textbf{Num.~Tokens} & \textbf{Time}      \\
		\hline
		EPU           & Manual          & articles      & 12009                 & N/A                  & \(\sim\) two years \\
		EPU~LDA       & Semi-Auto       & articles      & 40454                 & 1,000,000+           & several hours      \\
		EPU~WIG       & Automatic       & headlines     & 11934                 & 8802                 & \(\sim\) 15min     \\
		\hline
	\end{tabular}
	\caption{Comparison of the dataset among the three methods.
		WIG requires a much smaller dataset and runs faster.}
	\label{tab:comparison}
\end{table}

\begin{figure}[H]
	\centering
	\includegraphics[width=0.8\linewidth]{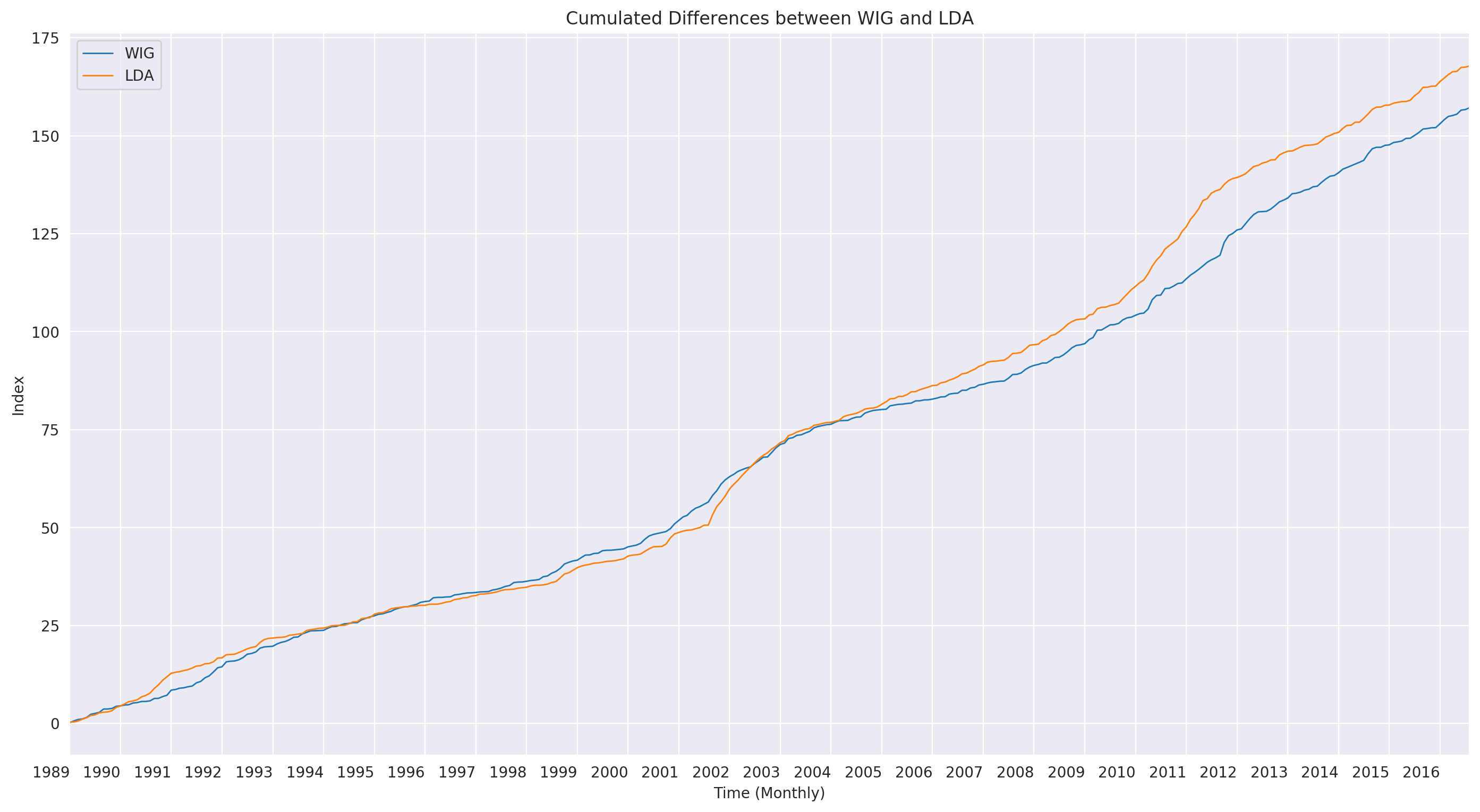}
	\caption{Cumulated difference between
		original EPU with EPU given by LDA and WIG.}
	\label{fig:cumsumdiff}
\end{figure}

\begin{table}[H]
	\centering
	\begin{tabular}{lccc}
		\hline
		\textbf{Correlation} & \textbf{Raw Series} & \textbf{Trend}  & \textbf{Cycle}  \\
		\hline
		\textbf{Pearson}     &                     &                 &                 \\
		EPU~LDA              & 0.7747              & 0.8679          & 0.7726          \\
		EPU~WIG              & \textbf{0.8023}     & \textbf{0.9093} & \textbf{0.7874} \\
		\hline
		\textbf{Spearman}    &                     &                 &                 \\
		EPU~LDA              & 0.7542              & 0.7666          & 0.7027          \\
		EPU~WIG              & \textbf{0.7749}     & \textbf{0.8631} & \textbf{0.7158} \\
	\end{tabular}
	\caption{Correlation between original EPU with the ones generated
		by WIG and LDA respectively, using Pearson's and Spearman's test.}
	\label{tab:correlation}
\end{table}

\end{document}